\documentclass[11pt]{article}
\usepackage[margin=1in]{geometry}
\usepackage{jheppub}
\usepackage[T1]{fontenc}
\usepackage[utf8]{inputenc}
\usepackage{amsmath,amssymb,mathtools,bm}
\usepackage{graphicx}
\usepackage{booktabs}
\usepackage{subcaption}
\usepackage{physics}
\usepackage{tikz-cd}
\usepackage{microtype}
\usepackage{enumitem}
\usepackage{bbm}
\usepackage{xcolor}
\usepackage{float}
\usepackage[nameinlink,noabbrev]{cleveref}

\newcommand{\ii}{\mathrm{i}}
\newcommand{\ee}{\mathrm{e}}
\newcommand{\mc}{m_{\mathrm{c}}}
\newcommand{\mDPT}{m_{\mathrm{DPT}}}

\newcommand{\Sigmeas}{\Sigma_{\mathrm{LOCC}}^{\mathrm{meas}}}
\newcommand{\Sigtomo}{\Sigma_{\mathrm{LOCC}}^{\mathrm{tomo}}}
\DeclareMathOperator*{\argmin}{arg\,min}

\title{Quantum Information Dynamics of QED$_2$ in Expanding de Sitter Universe}
\author[a,b]{Kazuki Ikeda}
\emailAdd{kazuki.ikeda@umb.edu}
\affiliation[a]{Department of Physics, University of Massachusetts Boston}
\affiliation[b]{Center for Nuclear Theory, Department of Physics and Astronomy, Stony Brook University}
\author[c]{Yaron Oz}
\affiliation[c]{School of Physics and Astronomy, Tel-Aviv University, Tel-Aviv 69978, Israel}
\emailAdd{yaronoz@tauex.tau.ac.il}
\date{\today}

\abstract{
We study QED$_2$ in de Sitter space as a minimal interacting gauge theory in which cosmological expansion directly competes with quantum dynamics. In cosmic time, the hopping redshifts as $1/a(t)$ while the electric term grows as $g^2 a(t)$, sweeping the spectrum through a moving narrow-gap region in the $(\tau,m)$ plane. Exact diagonalization shows that this defines a pseudo-critical line governing the loss of adiabaticity, excitation growth, and redshifted response. Using matrix-product states at a fixed mass, we separate the fixed-cutoff thermodynamic limit from the continuum extrapolation. The late-time dip survives in the infinite physical box size limit, and shifts to later $\tau$ as the lattice spacing goes to zero, with current data favoring $\tau_* \approx 3.1$, while the dip depth remains less controlled.
For Gibbs initial states, the same mechanism produces an irreversibility front in the relative entropy that tracks the pseudo-critical line and is detectable via LOCC-accessible observables. These results identify de Sitter QED$_2$ as a controlled setting for linking curved-space gauge dynamics, near-critical spectral structure, and operational irreversibility.
}

\begin{document}
\maketitle
\flushbottom

\section{Introduction}
Quantum field theory in curved spacetime provides the standard framework for particle production and nonequilibrium dynamics in expanding backgrounds \cite{Parker1969,BirrellDavies1982}. Recent work has pushed these questions toward controllable and genuinely nonperturbative settings, including cold-atom analogues of expanding Dirac fields, digital simulations of cosmological particle creation, interacting Dirac theories in Friedmann--Robertson--Walker backgrounds, and tensor-network studies of interacting fields in expanding geometries \cite{FulgadoClaudio2023,DiazMaceda2025,GongYang2025,FulgadoClaudio2024,BuddEtAl2026}. At the same time, there is a long continuum literature on Schwinger pair production and induced currents in de Sitter backgrounds \cite{Garriga1994,StahlStrobelXue2016}. What is still missing is a many-body picture of how an expanding geometry reshapes the spectrum and irreversibility of an interacting gauge theory with exact Gauss constraints.

The Schwinger model is especially well suited to this question. QED$_2$ is the simplest interacting gauge theory that retains confinement, pair production, and an exact local Gauss law, while remaining accessible to exact diagonalization, tensor networks, and present-day quantum-simulation platforms \cite{Schwinger1962,KogutSusskind1975,Banuls2020,Bauer2023,DiMeglio2024,Halimeh2025,PhysRevLett.109.125302,PhysRevA.98.032331}. Recent work has used the Schwinger model \emph{in flat spacetime} to study nonequilibrium dynamics, dynamical quantum phase transitions, entanglement structure, finite-temperature diagnostics, and quantum algorithms for ground and excited states \cite{deJong2022,Mueller2023,PRXQuantum.5.020315,Farrell2024,Kaikov2024,Guo2024,ArguelloCruz2024,PhysRevD.108.L091501,Grieninger2024FT,Ikeda:2024rzv,PhysRevD.108.074001,Florio2023,Florio2024,Barata:2024bzk,NicacioFalcao2025,GrieningerSavageZemlevskiy2026,Barata:2025rjb,Shao:2025ygy,Rogerson:2026jjt,Barata:2023jgd,Shaw2020quantumalgorithms,PhysRevResearch.2.023342,Barata:2024apg}. However nonequilibrium and perturbative real-time dynamics \emph{in de-Sitter space} is largely unexplored except a few cases \cite{BuddEtAl2026,GongYang2025,DiazMaceda2025,FulgadoClaudio2024,FulgadoClaudio2023}. Its importance here is not only numerical convenience. In de Sitter flat slicing, the cosmic-time Hamiltonian acquires a particularly clean structure: the hopping redshifts as $1/a(t)$ whereas the electric term grows as $g^2 a(t)$. The expansion therefore does not merely populate excitations; it continuously reweights delocalization against Coulomb energy (therefore the total energy does not conserve) and turns the real-time evolution into a controlled sweep through the interacting many-body spectrum.

This perspective is different from the one taken in most curved-spacetime studies. Free or self-interacting Dirac theories in expanding backgrounds primarily probe how the notion of particles is modified by the geometry \cite{FulgadoClaudio2024}. Very recently, Ref.~\cite{BuddEtAl2026} formulated the Schwinger model in general curved backgrounds and studied gravitational particle production and entanglement with tensor networks. Here we pursue a complementary question. We ask whether the de Sitter drive generates a moving pseudo-critical locus in the instantaneous spectrum, whether the associated late-time crossing survives once the physical volume is increased at fixed cutoff and the cutoff is subsequently reduced, and whether the same locus leaves an operational thermodynamic imprint that can be reconstructed from local data. To our knowledge, this combination of questions has not been addressed before in expanding QED$_2$ in de-Sitter cosmological setting.

A central issue is therefore what survives beyond the fixed-volume spectral map. In this paper the small-$N$ exact-diagonalization scans are used only to identify the moving pseudo-critical valley in the $(\tau,m)$ plane. The survival question is addressed separately by a two-step analysis: first the fixed-cutoff thermodynamic limit $\ell_{\rm phys}=N a_{\rm latt}\to\infty$ at fixed $a_{\rm latt}$, and only then the continuum extrapolation $a_{\rm latt}\to0$ of the thermodynamic branch limits. This separation is essential because a fixed-volume scan can reveal the spectral mechanism, but by itself it cannot decide whether the late-time dip persists at growing physical volume or under improving lattice resolution.

The main contributions of this paper are threefold. First, using exact diagonalization at fixed finite volume to resolve the full $(\tau,m)$ landscape, we show that expanding QED$_2$ naturally develops a moving narrow-gap valley whose locus defines a pseudo-critical line. Second, combining this finite-volume spectral map with matrix-product-state calculations at fixed $m=-1.5$, we separate the infrared question of what survives as $\ell_{\rm phys}=N a_{\rm latt}\to\infty$ from the continuum question of how the surviving crossing drifts as $a_{\rm latt}\to0$. The late-time dip time remains finite in the fixed-cutoff thermodynamic limit for every branch we study and moves monotonically to later $\tau$ as the cutoff is reduced, with the present data favoring $\tau_\ast^{(\infty,0)}\approx 3.1$; by contrast, the dip depth is less regular and is used only diagnostically. Third, for Gibbs initial states, we identify an irreversibility front in the relative entropy to the instantaneous Gibbs state, show that it tracks the same pseudo-critical line, and quantify how accurately it can be reconstructed by LOCC-accessible witnesses. The resulting picture is not simply that expansion distorts the spectrum, but that it generates a dynamically and thermodynamically visible threshold whose persistence can be tested separately in physical volume and lattice spacing.

This paper is organized as follows. \Cref{sec:model} derives the expanding-universe lattice Hamiltonian and introduces the observables used throughout. \Cref{sec:specdyn} presents the spectral flow and real-time diagnostics under ground-state initialization. \Cref{sec:largeN} then addresses the two-step large-system analysis, separating the fixed-cutoff thermodynamic extrapolation from the continuum drift up to $N=100$. \Cref{sec:front} formulates the entropy-production landscape, extracts the irreversibility front, and analyzes its LOCC reconstruction. We conclude in \Cref{sec:discussion,sec:conclusion}.

\section{From curved-space QED$_2$ to a qubit Hamiltonian}
\label{sec:model}

\subsection{Continuum QED$_2$ in an expanding universe}
We consider a spatially flat $(1+1)$-dimensional FLRW geometry,
\begin{equation}
  \dd s^2 = a^2(\eta)\left(-\dd\eta^2+\dd x^2\right)
  = -\dd t^2 + a^2(t)\,\dd x^2,
  \qquad \dd t = a(\eta)\,\dd\eta,
  \label{eq:FLRW_metric}
\end{equation}
where $\eta$ is conformal time and $t$ is cosmic time. Throughout the dynamical analysis we focus on de Sitter flat slicing,
\begin{equation}
  a(t)=\ee^{t/L},
  \qquad
  \tau \equiv \frac{t}{L},
  \qquad
  H(t)\equiv \frac{\dot a}{a}=\frac{1}{L},
  \label{eq:dS_flat_slicing}
\end{equation}
so that the dimensionless time variable $\tau$ is the natural control parameter in the figures.

We start from the curved-space QED$_2$ action
\begin{equation}
  S = \int \dd^2x\,\sqrt{-g}\,
  \bar\Psi\left(\ii\gamma^\mu(\nabla_\mu-\ii A_\mu)-m\right)\Psi
  + \mu\int \dd^2x\,\sqrt{-g}\,\Psi^\dagger\Psi
  - \frac{1}{4g^2}\int \dd^2x\,\sqrt{-g}\,F_{\mu\nu}F^{\mu\nu},
  \label{eq:QED2_action}
\end{equation}
where $\nabla_\mu$ contains the spin connection and $F_{\mu\nu}=\partial_\mu A_\nu-\partial_\nu A_\mu$. The time dependence relevant for our later lattice Hamiltonian is most transparent in conformal time. In the metric \eqref{eq:FLRW_metric}, the Maxwell term reduces to
\begin{equation}
  -\frac{1}{4g^2}\sqrt{-g}\,F_{\mu\nu}F^{\mu\nu}
  = \frac{1}{2g^2 a^2(\eta)}F_{\eta x}^2,
  \qquad F_{\eta x}=\partial_\eta A_x-\partial_x A_\eta,
  \label{eq:maxwell_conformal}
\end{equation}
so the canonical momentum conjugate to $A_x$ is
\begin{equation}
  \Pi_x(\eta,x)=\frac{\partial\mathcal{L}}{\partial(\partial_\eta A_x)}
  =\frac{1}{g^2 a^2(\eta)}F_{\eta x}.
  \label{eq:Pi_def}
\end{equation}
The corresponding gauge contribution to the conformal-time Hamiltonian is therefore
\begin{equation}
  H_{E,\eta}(\eta)
  =\int \dd x\left[\frac{g^2 a^2(\eta)}{2}\,\Pi_x^2 - A_\eta\,\partial_x\Pi_x\right],
  \label{eq:HE_conformal}
\end{equation}
with Gauss' law imposed by $A_\eta$:
\begin{equation}
  \partial_x\Pi_x = \rho(\eta,x).
  \label{eq:gauss_continuum}
\end{equation}

The passage from conformal to cosmic time is equally important for the fermion sector. If the Schr\"odinger equation is written as
\begin{equation}
  \ii\partial_\eta \ket{\Psi(\eta)} = H_\eta(\eta)\ket{\Psi(\eta)},
\end{equation}
then using $\partial_\eta = a(t)\partial_t$ yields
\begin{equation}
  \ii\partial_t \ket{\Psi(t)} = H_t(t)\ket{\Psi(t)},
  \qquad
  H_t(t)=\frac{1}{a(t)}H_\eta\bigl(\eta(t)\bigr).
  \label{eq:Heta_to_Ht}
\end{equation}
Thus the cosmic-time Hamiltonian inherits distinct scale-factor dependencies for different terms:
\begin{align}
  &\text{kinetic term:} && J(t)\propto \frac{1}{a(t)},
  \label{eq:kinetic_redshift}\\
  &\text{mass and chemical potential:} && m,\mu \ \text{remain time independent},\\
  &\text{spin connection:} && \frac{a'(\eta)}{2a(\eta)} \to \frac{H(t)}{2},\\
  &\text{electric energy:} && \frac{g^2 a^2(\eta)}{2}\Pi_x^2 \to \frac{g^2 a(t)}{2}\Pi_x^2.
  \label{eq:geff_cosmic}
\end{align}
For constant microscopic gauge coupling $g$ in the covariant action \eqref{eq:QED2_action}, the physical electric term therefore carries the effective prefactor
\begin{equation}
  g_{\mathrm{eff}}^2(t)=g^2 a(t).
  \label{eq:geff}
\end{equation}
This simple observation underlies all of the nonequilibrium phenomena discussed below: expansion simultaneously suppresses hopping and amplifies the electric energy.

\subsection{Lattice discretization and Gauss-law elimination}
We discretize the comoving coordinate as $x_n=n\,a_{\rm lat}$ with $n=1,\dots,N$ and use the Kogut--Susskind staggered formulation \cite{KogutSusskind1975}. The lattice degrees of freedom are a one-component fermion $\chi_n$ on each site and a compact $U(1)$ link operator $U_n$ on the link $(n,n+1)$ with conjugate electric field $L_n$, satisfying
\begin{equation}
  [L_n,U_m]=\delta_{nm}\,U_n.
\end{equation}
In cosmic time the gauge-invariant lattice Hamiltonian takes the form
\begin{align}
  H_{\rm lat}(t)
  &=
  -\frac{\ii}{2a_{\rm lat}a(t)}\sum_{n=1}^{N-1}
  \left(\chi_n^\dagger U_n \chi_{n+1}-\chi_{n+1}^\dagger U_n^\dagger \chi_n\right)
  \nonumber\\
  &\quad
  +\sum_{n=1}^{N}\left(m(-1)^n-\mu+\frac{H(t)}{2}\right)\chi_n^\dagger\chi_n
  +\frac{g^2 a(t)a_{\rm lat}}{2}\sum_{n=1}^{N-1}L_n^2
  +(\text{constant}).
  \label{eq:lattice_QED2_FLRW}
\end{align}
The term proportional to $H(t)$ is the spin-connection contribution in the unrescaled cosmic-time fermion variable. It is absent in formulations such as Ref.~\cite{BuddEtAl2026}, where the standard Weyl rescaling of the Dirac field, $\Psi=a^{-1/2}\psi$, is made before discretization; in that convention the spin connection is absorbed into the field normalization. Equivalently, in the Hamiltonian above it can be viewed as the chemical-potential shift $\mu\to\mu-H(t)/2$. In the fixed charge sector used below this contribution is a sector constant, while in the Gibbs-state calculations it is kept consistently as part of the chosen chemical-potential convention.

Gauss' law at each site is
\begin{equation}
  G_n \equiv L_n-L_{n-1}-q_n \approx 0,
  \qquad n=1,\dots,N,
  \label{eq:Gauss_lattice}
\end{equation}
with the staggered charge
\begin{equation}
  q_n = \chi_n^\dagger\chi_n - \frac{1-(-1)^n}{2}.
  \label{eq:staggered_charge}
\end{equation}
For open boundaries and left boundary electric field $L_0=E_0$, Gauss' law can be solved exactly as
\begin{equation}
  L_n = E_0 + \sum_{k=1}^{n} q_k,
  \qquad n=1,\dots,N-1.
  \label{eq:L_solution_open}
\end{equation}
The last Gauss-law equation fixes the right boundary flux,
\[
  L_N=E_0+Q_{\rm tot},\qquad Q_{\rm tot}\equiv\sum_{k=1}^{N}q_k .
\]
Thus a fixed charge sector is equivalently a fixed right electric-field boundary condition. In the numerical results below we use the neutral sector $Q_{\rm tot}=0$; with $E_0=0$ this gives $L_0=L_N=0$. Choosing axial gauge $U_n=\mathbbm{1}$ then removes the link variables explicitly, so the gauge sector survives only through the nonlocal electric-energy term built from the cumulative charge.

\subsection{Jordan--Wigner map and the full qubit Hamiltonian}
To obtain a qubit Hamiltonian we apply the Jordan--Wigner transformation \cite{JordanWigner1928},
\begin{equation}
  \chi_n = \frac{X_n-\ii Y_n}{2}\prod_{j<n}(-\ii Z_j),
  \qquad
  \chi_n^\dagger\chi_n = \frac{1+Z_n}{2}.
  \label{eq:JW_map}
\end{equation}
The staggered charge becomes
\begin{equation}
  q_n = \frac{Z_n+(-1)^n}{2},
  \label{eq:q_in_Z}
\end{equation}
and the hopping term maps to the nearest-neighbor $XY$ coupling,
\begin{equation}
  -\frac{\ii}{2a_{\rm lat}a(t)}
  \left(\chi_n^\dagger\chi_{n+1}-\chi_{n+1}^\dagger\chi_n\right)
  \longrightarrow
  \frac{1}{4a_{\rm lat}a(t)}(X_nX_{n+1}+Y_nY_{n+1}).
  \label{eq:hop_to_XY}
\end{equation}
Putting everything together, the full qubit Hamiltonian reads
\begin{equation}
  \begin{aligned}
  H(t)
  &=
  \sum_{n=1}^{N-1}\frac{1}{4a_{\rm lat}a(t)}\left(X_nX_{n+1}+Y_nY_{n+1}\right)
  +\sum_{n=1}^{N}\frac{m}{2}(-1)^n Z_n
  +\left(-\frac{\mu}{2}+\frac{H(t)}{4}\right)\sum_{n=1}^{N} Z_n
  \\
  &\quad
  +\frac{g^2 a(t)a_{\rm lat}}{2}\sum_{n=1}^{N-1}
  \left(E_0+\sum_{k=1}^{n}\frac{Z_k+(-1)^k}{2}\right)^2
  +(\text{constant}).
  \end{aligned}
  \label{eq:qubit_QED2_FLRW_full}
\end{equation}

\subsection{Competing diagonal energies and the moving diabatic crossing}
\label{sec:diabatic_crossing}
A useful reorganization of \eqref{eq:qubit_QED2_FLRW_full} is
\begin{equation}
  H(\tau,m)=J(\tau)\,K + m\,M + h_{\rm FLRW}(\tau)\,Q + \lambda(\tau)\,W + (\text{constant}),
  \label{eq:H_decomp}
\end{equation}
where
\begin{align}
  J(\tau) &= \frac{1}{4a_{\rm lat}a(\tau)},
  &\lambda(\tau) &= \frac{g^2 a(\tau)a_{\rm lat}}{2},
  &h_{\rm FLRW}(\tau) &= -\frac{\mu}{2}+\frac{H(\tau)}{4},
  \nonumber\\
  K &= \sum_{n=1}^{N-1}\left(X_nX_{n+1}+Y_nY_{n+1}\right),
  &M &= \frac{1}{2}\sum_{n=1}^{N}(-1)^n Z_n,
  &Q &= \sum_{n=1}^{N} Z_n,
  \nonumber\\
  W &= \sum_{n=1}^{N-1}\left(E_0+\sum_{k=1}^{n}\frac{Z_k+(-1)^k}{2}\right)^2.
\end{align}
In the computational $Z$-basis, the operator
\begin{equation}
  H_{\rm diag}(\tau,m)\equiv mM+h_{\rm FLRW}(\tau)Q+\lambda(\tau)W
\end{equation}
is diagonal, while the kinetic term $J(\tau)K$ is the only off-diagonal piece. For a basis configuration $\mathbf z=(z_1,\dots,z_N)$ with $z_n=\pm 1$, the corresponding diabatic energy is
\begin{equation}
  \mathcal E_{\mathbf z}(\tau,m)
  = m\,M[\mathbf z] + h_{\rm FLRW}(\tau)\,Q[\mathbf z] + \lambda(\tau)\,W[\mathbf z],
  \label{eq:diabatic_energy}
\end{equation}
with
\begin{equation}
  M[\mathbf z]=\frac{1}{2}\sum_{n=1}^{N}(-1)^n z_n,
  \qquad
  Q[\mathbf z]=\sum_{n=1}^{N}z_n,
\end{equation}
and
\begin{equation}
  W[\mathbf z]=\sum_{n=1}^{N-1}\left(E_0+\sum_{k=1}^{n}\frac{z_k+(-1)^k}{2}\right)^2.
\end{equation}
Thus, in the late-time regime where $J(\tau)\ll \lambda(\tau)$, the low-lying spectrum is controlled by crossings between these diabatic charge configurations. For two configurations $A$ and $B$, the crossing condition $\mathcal E_A=\mathcal E_B$ gives
\begin{equation}
  m_c^{AB}(\tau)
  =
  -\frac{h_{\rm FLRW}(\tau)\,\Delta Q^{AB}+\lambda(\tau)\,\Delta W^{AB}}{\Delta M^{AB}},
  \qquad
  \Delta O^{AB}\equiv O[B]-O[A].
  \label{eq:mc_AB}
\end{equation}
This makes the origin of the drift transparent: because $\lambda(\tau)\propto a(\tau)$, the location of each diabatic crossing moves with the expansion, and for the branch pair relevant to the observed valley the sign of $\Delta W^{AB}/\Delta M^{AB}$ drives $m_c^{AB}(\tau)$ toward more negative values. The hopping term does not create the drift; it only converts the drifting diabatic crossing into an avoided crossing. If the direct matrix element between the two branches is nonzero, the minimal gap is
\begin{equation}
  \Delta_{\min}^{AB}(\tau)\simeq 2\abs{\Gamma_{AB}(\tau)},
  \qquad
  \Gamma_{AB}(\tau)=J(\tau)\mel{A}{K}{B},
  \label{eq:gap_AB_direct}
\end{equation}
and the same statement holds with the appropriate higher-order matrix element when the direct coupling vanishes. Since $J(\tau)\propto a^{-1}(\tau)$, later avoided crossings are parametrically narrower. This is the microscopic reason the expanding Hamiltonian develops a moving narrow-gap valley rather than a static critical point.

This is the basic microscopic input for the rest of the paper. In the flat limit $a(t)=1$ and $H(t)=0$, \eqref{eq:qubit_QED2_FLRW_full} reduces to the standard lattice Schwinger-model Hamiltonian. In de Sitter flat slicing \eqref{eq:dS_flat_slicing}, by contrast, one has
\begin{equation}
  J(t)=\frac{\ee^{-t/L}}{4a_{\rm lat}},
  \qquad
  H(t)=\frac{1}{L},
  \qquad
  g_{\rm eff}^2(t)=g^2\ee^{t/L},
\end{equation}
so the expansion itself produces a built-in sweep through coupling space.

\subsection{Protocols and observables}
We will use two complementary initializations. For the coherent real-time diagnostics of \Cref{sec:specdyn}, the system is initialized in the instantaneous ground state at $t=0$ within a fixed charge sector and evolved under the explicitly time-dependent Hamiltonian \eqref{eq:qubit_QED2_FLRW_full}. For the thermodynamic and operational analysis of \Cref{sec:front}, the initial state is instead the Gibbs state of the initial Hamiltonian,
\begin{equation}
  \rho_0 = \rho_\beta(H(0)) = \frac{\ee^{-\beta H(0)}}{\Tr\,\ee^{-\beta H(0)}},
  \label{eq:initial_gibbs}
\end{equation}
which is then evolved by the time-ordered unitary
\begin{equation}
  U(\tau,0)=\mathcal{T}\exp\!\left[-\ii\int_0^{\tau}L H(\tau')\,\dd\tau'\right].
  \label{eq:time_ordered_U}
\end{equation}

The observables common to the two protocols are the instantaneous many-body gap
\begin{equation}
  \Delta(\tau,m)=E_1(\tau,m)-E_0(\tau,m),
  \label{eq:gap_def}
\end{equation}
and the pseudo-critical line
\begin{equation}
  \mc(\tau)=\argmin_m\Delta(\tau,m),
  \label{eq:mcdef}
\end{equation}
which identifies the mass value where the instantaneous spectrum is narrowest at fixed $\tau$.

For the ground-state protocol we also monitor the instantaneous ground-state fidelity
\begin{equation}
  F_{\rm GS}(t)=\abs{\langle{\phi_0(t)|\psi(t)}\rangle}^2,
  \label{eq:Fgs}
\end{equation}
the excitation energy density
\begin{equation}
  \epsilon_{\rm exc}(t)=\frac{\mel{\psi(t)}{H(t)}{\psi(t)}-E_0(t)}{N},
  \label{eq:eps_exc}
\end{equation}
and the momentum-resolved charge response. Using the staggered charge \eqref{eq:q_in_Z}, we define the connected structure factor
\begin{equation}
  S_q(k,t)=\frac{1}{N}\sum_{r,s=1}^{N}\ee^{\ii k(r-s)}\left(\ev{q_r q_s}_t-\ev{q_r}_t\ev{q_s}_t\right),
  \label{eq:Sq_def}
\end{equation}
For the heat map in \Cref{fig:paper_summary}(d) we plot the change relative to the initial state,
\[
  \Delta S_q(k,\tau)\equiv S_q(k,\tau)-S_q(k,0),
\]
so that the color scale isolates the expansion-induced redistribution of charge correlations. The peak of $S_q$ (equivalently, of the plotted response after this subtraction) in comoving momentum space can be compared with the physical redshift law $p(t)=k/a(t)$.

For the Gibbs-state protocol we define the entropy production
\begin{equation}
  \Sigma(\tau,m;\beta)
  = D\!\left(\rho(\tau)\,\Vert\,\rho_\beta(H(\tau))\right) \ge 0,
  \label{eq:sigma}
\end{equation}
where $D(\rho\Vert\sigma)=\Tr[\rho(\log\rho-\log\sigma)]$ is the quantum relative entropy \cite{Umegaki1962,Vedral2002}. The LOCC-accessible witnesses used later are
\begin{equation}
  \Sigmeas = D_{\rm cl}\!\left(p_{AB}\,\Vert\,q_{AB}\right),
  \qquad
  \Sigtomo = D\!\left(\rho_{AB}\,\Vert\,\sigma_{AB}\right),
  \label{eq:locc_witnesses}
\end{equation}
Here $A$ and $B$ are end blocks of size $\ell$ controlled by Alice and Bob. For $\Sigmeas$, Alice and Bob perform local projective measurements in the staggered-charge, equivalently computational $Z$, basis and then compare their outcomes by classical communication. With $\Pi^A_{\mathbf z_A}=|\mathbf z_A\rangle\langle\mathbf z_A|$ and $\Pi^B_{\mathbf z_B}=|\mathbf z_B\rangle\langle\mathbf z_B|$,
\[
  p_{AB}(\mathbf z_A,\mathbf z_B)=
  \Tr\!\left[(\Pi^A_{\mathbf z_A}\otimes\Pi^B_{\mathbf z_B})\,\rho_{AB}(\tau)\right],\qquad
  q_{AB}(\mathbf z_A,\mathbf z_B)=
  \Tr\!\left[(\Pi^A_{\mathbf z_A}\otimes\Pi^B_{\mathbf z_B})\,\sigma_{AB}(\tau)\right],
\]
where $\rho_{AB}(\tau)=\Tr_{\overline{AB}}\rho(\tau)$ and $\sigma_{AB}(\tau)=\Tr_{\overline{AB}}\rho_\beta(H(\tau))$. The tomography witness $\Sigtomo$ uses the same reduced density matrices before this final local measurement channel. By monotonicity of relative entropy under partial trace and measurement,
\begin{equation}
  \Sigma \ge \Sigtomo \ge \Sigmeas.
  \label{eq:locc_bound}
\end{equation}
This provides a direct operational bridge between global nonequilibrium dynamics and local access by distant observers.

\section{Spectral flow and coherent expansion dynamics}
\label{sec:specdyn}

We first analyze the coherent dynamical sweep generated by the expanding Hamiltonian \eqref{eq:qubit_QED2_FLRW_full} under ground-state initialization. Unless stated otherwise we work in a fixed charge sector with open boundaries and use the representative parameters
\begin{equation}
  a_{\rm lat}=1,
  \qquad
  g=1,
  \qquad
  \mu=0,
  \qquad
  E_0=0.
  \label{eq:base_params}
\end{equation}
All dimensionful numerical quantities are quoted in units of $g$. The charge sector is the neutral sector $Q_{\rm tot}=0$ specified above, so the right boundary field is $L_N=0$ for $E_0=0$. The instantaneous spectral diagnostics in \Cref{fig:paper_summary}(a) and \Cref{fig:largeN_fixedmass} are plotted as functions of $\tau=t/L$ and do not require an additional choice of de Sitter radius in this sector: changing $L$ rescales the physical crossing time as $t_c=L\tau_c$, whereas the real-time adiabaticity curves in \Cref{fig:paper_summary}(b,c) display the explicit dependence on the sweep rate $1/L$. The real-time propagation is implemented with exact diagonalization using a midpoint piecewise-constant approximation for the time-ordered exponential.

\subsection{Effective two-level sweep and nonadiabatic scaling}
The decomposition \eqref{eq:H_decomp} implies a natural effective description near the lowest valley. Once a particular pair of diabatic branches $A,B$ dominates, we may project the many-body Hamiltonian onto their span. To leading order in the redshifted hopping,
\begin{equation}
  H_{\rm eff}^{AB}(\tau,m)
  = \varepsilon_+^{AB}(\tau,m)\,\mathbbm{1}
  +\frac{\delta_{AB}(\tau,m)}{2}\sigma_z
  +\Gamma_{AB}(\tau)\sigma_x,
  \label{eq:H_eff_AB}
\end{equation}
with
\begin{equation}
  \delta_{AB}(\tau,m)=\Delta M^{AB}\!\left[m-m_c^{AB}(\tau)\right]+\mathcal O(J^2),
  \qquad
  \Gamma_{AB}(\tau)=J(\tau)\mel{A}{K}{B},
  \label{eq:deltaAB}
\end{equation}
and instantaneous gap
\begin{equation}
  \Delta_{AB}(\tau,m)=\sqrt{\delta_{AB}^2(\tau,m)+4\abs{\Gamma_{AB}(\tau)}^2}.
  \label{eq:gap_eff}
\end{equation}
Equation \eqref{eq:gap_eff} shows that the mass width of the valley at fixed time scales as
\begin{equation}
  \delta m_{\rm valley}^{AB}(\tau)\sim \frac{2\abs{\Gamma_{AB}(\tau)}}{\abs{\Delta M^{AB}}}
  \propto \frac{1}{a(\tau)},
  \label{eq:valley_width_scaling}
\end{equation}
again up to the corresponding higher-order replacement of $\Gamma_{AB}$ when the direct matrix element vanishes. The expansion therefore has a double effect: it translates the crossing line through \eqref{eq:mc_AB} and simultaneously narrows it through \eqref{eq:valley_width_scaling}.

For a fixed mass $m$, the real-time dynamics is a sweep through $\delta_{AB}(t,m)=0$ at some crossing time $t_c$. The relevant detuning rate is
\begin{equation}
  v_{AB}(t_c)\equiv \abs{\dot\delta_{AB}(t_c,m)}
  = \abs{\Delta M^{AB}\,\dot m_c^{AB}(t_c)},
  \label{eq:vAB}
\end{equation}
which in the late-time regime is dominated by the electric term and scales as
\begin{equation}
  v_{AB}(t_c)\simeq \frac{g^2 a_{\rm lat}}{2L}\,a(t_c)\,\abs{\Delta W^{AB}}.
  \label{eq:vAB_late}
\end{equation}
The associated excitation probability is therefore of Landau--Zener form,
\begin{equation}
  P_{\rm ex}^{AB}(m)\approx
  \exp\!\left[-\frac{2\pi \abs{\Gamma_{AB}(t_c)}^2}{v_{AB}(t_c)}\right],
  \label{eq:LZ_form}
\end{equation}
so the exponent scales schematically as $L/a^3(t_c)$ once the late-time asymptotics $J\sim a^{-1}$ and $v_{AB}\sim a/L$ are inserted. This immediately explains two of the numerical trends below: later crossings are more non-adiabatic, and smaller de Sitter radii $L$ produce stronger fidelity loss and larger excitation energy.

\subsection{Pseudo-critical line from the instantaneous gap landscape}
The gap landscape $\Delta(\tau,m)$ provides the most economical way to visualize how the expansion reshuffles the spectrum. A representative example is shown in \Cref{fig:paper_summary}(a). The white curve traces the pseudo-critical line $\mc(\tau)$ and moves steadily toward more negative mass as the universe expands. In the effective description \eqref{eq:gap_eff}, this is exactly the drift of the underlying diabatic crossing line $m_c^{AB}(\tau)$, while the narrowness of the valley is controlled by the decaying mixing amplitude $\Gamma_{AB}(\tau)$. The observed pseudo-critical line is thus the finite-volume avoided-crossing image of the competition encoded in \eqref{eq:qubit_QED2_FLRW_full}: the kinetic scale is redshifted as $1/a(t)$, while the Coulomb scale grows as $g^2 a(t)$.

A representative crossing point lies near $m_\ast\simeq -1.5$ and $\tau_c\simeq 1.8$. This point is not special in principle; it is simply a convenient location where the instantaneous gap becomes narrow enough that coherent evolution can no longer follow the ground state adiabatically. In this sense the pseudo-critical line serves as the finite-volume remnant of a critical manifold in the $(\tau,m)$ plane.

\subsection{Breakdown of adiabatic following and excitation growth}
The dynamical consequences of crossing the narrow-gap valley are summarized in \Cref{fig:paper_summary}(b),(c). Panel (b) shows the non-adiabaticity $1-F_{\rm GS}(\tau)$ for several curvature radii $L$. The curves remain close to zero at early times and then rise abruptly near the same time window where the gap is smallest. This is precisely the behavior expected from the sweep estimate \eqref{eq:LZ_form}: decreasing $L$ increases the detuning rate $v_{AB}$ and therefore suppresses the Landau--Zener exponent, while the flat limit $L\to\infty$ removes the sweep altogether and remains adiabatic up to numerical precision.

Panel (c) displays the excitation energy density $\epsilon_{\rm exc}(\tau)$ for the same representative parameters. Its onset follows the loss of fidelity and marks the injection of energy above the instantaneous ground state. The behavior is again curvature ordered: stronger expansion produces a larger final excitation density. Together, panels (b) and (c) show that the pseudo-critical line extracted from the instantaneous gap is not a merely geometric feature of the spectrum. It is the region where adiabatic tracking fails and the expansion produces real many-body excitations.

\subsection{Momentum-resolved response and cosmological redshift}
\begin{figure*}[t]
  \centering
  \includegraphics[width=\textwidth]{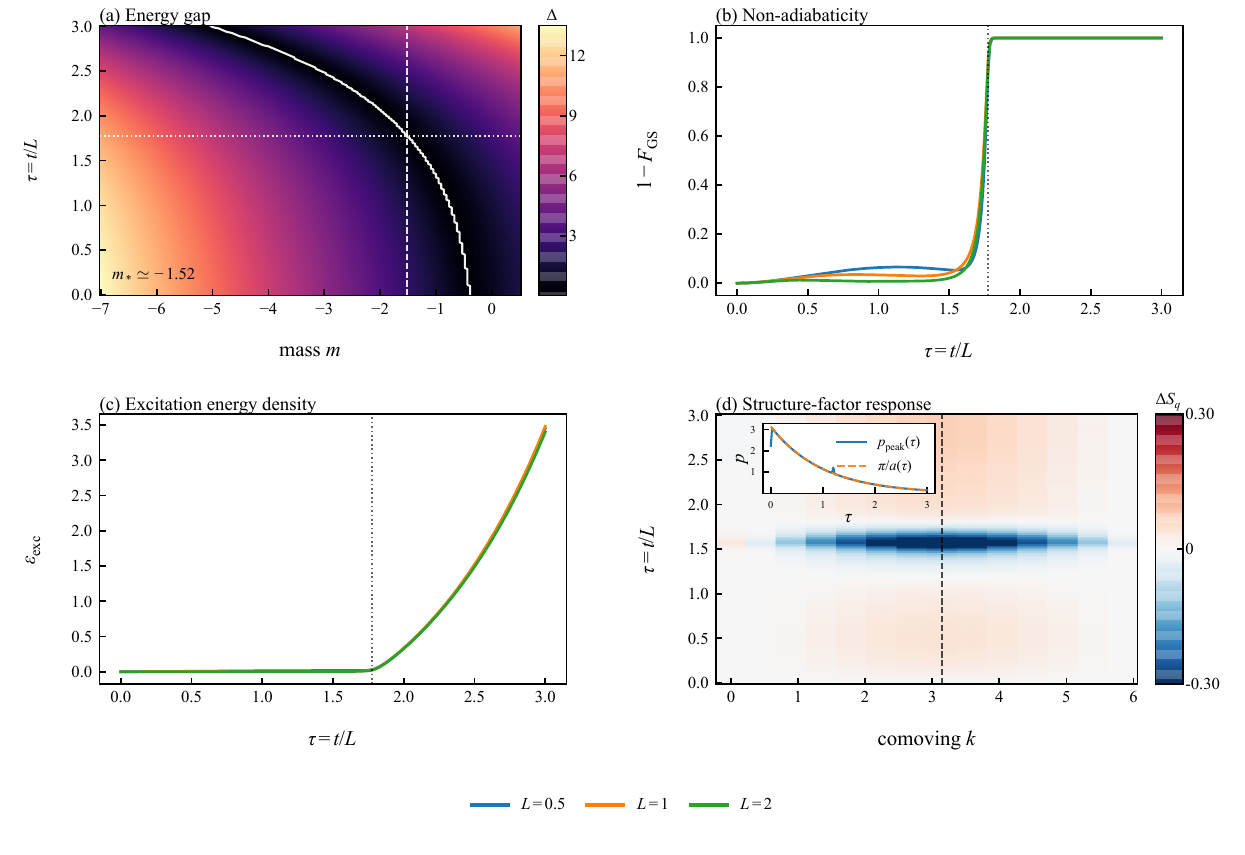}
  \caption{\textbf{Representative exact-diagonalization diagnostics of the expansion-driven sweep.}
  \textbf{(a)} Instantaneous gap landscape $\Delta(\tau,m)$ with pseudo-critical line $\mc(\tau)=\argmin_m\Delta(\tau,m)$ (white). Dashed lines indicate a representative crossing point $(\tau_c,m_\ast)$. \textbf{(b)} Non-adiabaticity $1-F_{\rm GS}(\tau)$ for several curvature radii $L$ at fixed $m_\ast$. \textbf{(c)} Excitation energy density $\epsilon_{\rm exc}(\tau)$ for the same parameters. \textbf{(d)} Structure-factor response $\Delta S_q(k,\tau)$, together with a check of the redshift law $p_{\rm peak}(\tau)\simeq\pi/a(\tau)$ in the inset. The main panel in (d) is plotted in comoving momentum, while the inset converts the peak to physical momentum. Panel (a) is an instantaneous spectral plot in the fixed neutral sector; the explicit dependence on $L$ enters the real-time sweep shown in panels (b) and (c). The data illustrate how the expansion-induced deformation of the spectrum translates into loss of adiabatic following, energy injection, and a physically redshifted response.}
  \label{fig:paper_summary}
\end{figure*}

The structure-factor response in \Cref{fig:paper_summary}(d) makes the cosmological interpretation particularly transparent. The dominant response stays near the staggered mode $k\simeq\pi$ in comoving momentum space, as expected from the staggered charge operator. The main heat map uses comoving momentum $k$, so the cosmological redshift is not expected to appear as a diagonal drift of the dominant band in that panel. The redshift appears only after converting the comoving peak to the physical momentum $p_{\rm peak}(\tau)=k_{\rm peak}(\tau)/a(\tau)$, which is what is shown in the inset. However, the inset demonstrates that the corresponding physical peak momentum obeys the redshift law
\begin{equation}
  p_{\rm peak}(\tau)\approx \frac{\pi}{a(\tau)} = \pi\ee^{-\tau}.
\end{equation}
Thus the expanding-universe lattice Hamiltonian simultaneously shows two distinct signatures: a moving pseudo-critical line in parameter space and the redshift of physical momenta in dynamical response functions.

\section{Fixed-cutoff thermodynamic limit and continuum drift}
\label{sec:largeN}

The exact-diagonalization landscape of \Cref{fig:paper_summary}(a) is intentionally shown at a small fixed volume so that the full $(\tau,m)$ structure can be resolved. The fate of the late-time dip under increasing physical size and decreasing lattice spacing is a separate question. To disentangle these effects, we perform a complementary matrix-product-state calculation at fixed representative mass
\begin{equation}
  m=-1.5,
  \qquad
  g=1,
  \qquad
  \mu=0,
  \qquad
  E_0=0,
  \label{eq:largeN_params}
\end{equation}
and separate the two limits
\begin{equation}
  \tau_\ast^{(\infty)}(a_{\rm latt})
  = \lim_{\ell_{\rm phys}\to\infty}\tau_\ast(a_{\rm latt},\ell_{\rm phys}),
  \qquad
  \tau_\ast^{(\infty,0)}
  = \lim_{a_{\rm latt}\to 0}\tau_\ast^{(\infty)}(a_{\rm latt}),
  \label{eq:two_limits_tau}
\end{equation}
where $\ell_{\rm phys}\equiv N a_{\rm latt}$ denotes the physical box size. To compare different cutoff branches at comparable physical volumes, we use the common ladder
\begin{equation}
  a_{\rm latt}\in\{1.0,0.8,0.6,0.48\},
  \qquad
  \ell_{\rm phys}\approx\{16,24,32,40,48\},
  \label{eq:thermo_grid}
\end{equation}
with the nearest integer $N$ on each branch and $N\le 100$. For each pair $(a_{\rm latt},\ell_{\rm phys})$ we compute the two lowest instantaneous eigenvalues of \eqref{eq:qubit_QED2_FLRW_full}, define the instantaneous gap
\begin{equation}
  \Delta(\tau;a_{\rm latt},\ell_{\rm phys})=E_1(\tau)-E_0(\tau),
\end{equation}
and locate the late-time dip by
\begin{equation}
  \tau_\ast(a_{\rm latt},\ell_{\rm phys})=\argmin_\tau \Delta(\tau;a_{\rm latt},\ell_{\rm phys}),
  \qquad
  \Delta_\ast(a_{\rm latt},\ell_{\rm phys})=\min_\tau \Delta(\tau;a_{\rm latt},\ell_{\rm phys}).
  \label{eq:taustar_def}
\end{equation}
Each run first locates the dip on a global time grid and then refines it in a narrower local window. All four cutoff branches have positive refined gaps and no edge-locked minima.

\begin{figure}[h]
  \centering
  \includegraphics[width=\textwidth]{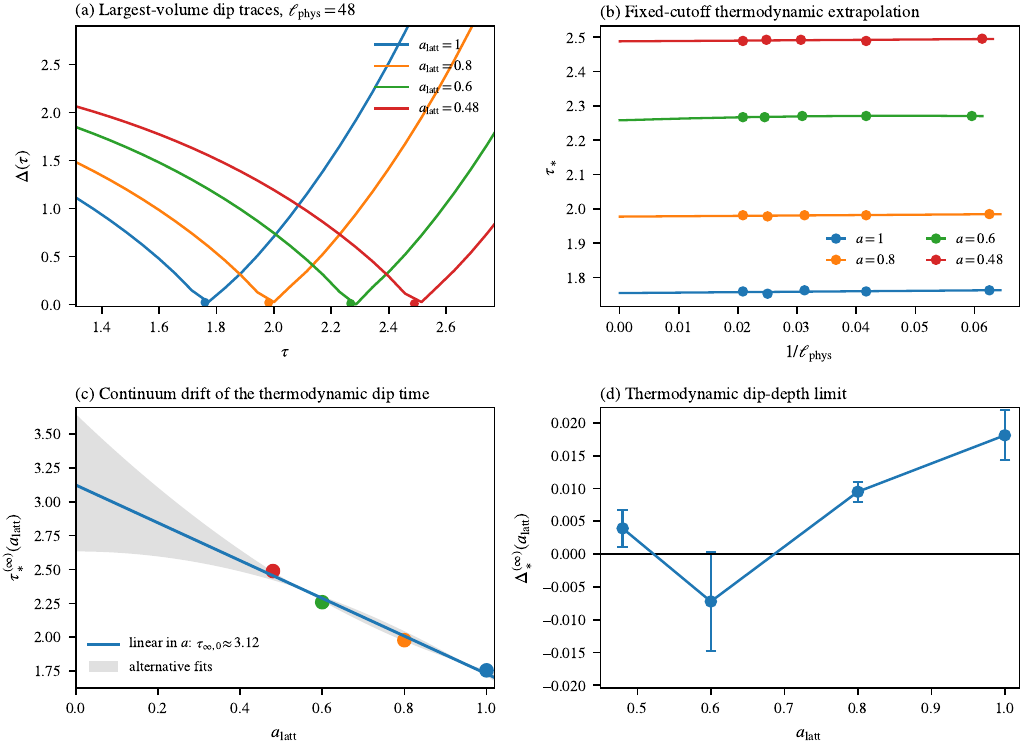}
  \caption{\textbf{Fixed-cutoff thermodynamic limit and continuum drift from matrix-product states.}
  \textbf{(a)} Largest-volume gap traces $\Delta(\tau)$ at $\ell_{\rm phys}=48$ for $a_{\rm latt}=1.0,0.8,0.6,0.48$, with markers at the refined dip times. \textbf{(b)} Fixed-cutoff thermodynamic extrapolation of $\tau_\ast(a_{\rm latt},\ell_{\rm phys})$ versus $1/\ell_{\rm phys}$ for the same four branches. The common physical-length ladder is $\ell_{\rm phys}\approx 16,24,32,40,48$. \textbf{(c)} Continuum drift of the thermodynamic dip time $\tau_\ast^{(\infty)}(a_{\rm latt})$. The solid line is the fit linear in $a_{\rm latt}$, giving $\tau_\ast^{(\infty,0)}\approx 3.12$, while the dashed line is the $a_{\rm latt}^2$ check, giving $2.63$. The shadow indicates fitting errors. \textbf{(d)} Thermodynamic-limit dip depth $\Delta_\ast^{(\infty)}(a_{\rm latt})$, shown only as a diagnostic. The dip time survives the fixed-cutoff thermodynamic limit and drifts to later times as the continuum is approached, while the dip depth itself remains less regular. Error bars in (d) denote numerical and extrapolation uncertainties rather than statistical sampling errors. Error bars in the other panels are not shown because they are negligible. For each fixed $a_{\rm latt}$, the uncertainty of $\Delta_{\ast}$ is estimated from the difference between the refined and coarse dip-depth extractions, with a small positive floor to avoid spuriously vanishing bars. These pointwise uncertainties are then propagated through the thermodynamic extrapolation in $1/\ell_{\rm phys}$, and the final error bar on $\Delta_{\ast}^{(\infty)}(a_{\rm latt})$ is obtained by combining in quadrature the weighted-fit intercept error and a model-spread estimate given by one half of the difference between the linear and quadratic extrapolated intercepts.}
  \label{fig:largeN_fixedmass}
\end{figure}

The results are summarized in \Cref{fig:largeN_fixedmass}. 
Panel (a) shows the largest-volume gap traces $\ell_{\rm phys}=48$ for the four cutoff branches. The late-time dip is present in every branch and moves monotonically to later $\tau$ as the lattice spacing is reduced. Panel (b) then addresses the first limit in \eqref{eq:two_limits_tau}. At fixed $a_{\rm latt}$, the dependence of $\tau_\ast(a_{\rm latt},\ell_{\rm phys})$ on $1/\ell_{\rm phys}$ is weak and smooth, so the thermodynamic limit can be taken branch by branch. The extrapolated dip times are
\begin{equation}
  \tau_\ast^{(\infty)}(a_{\rm latt})
  \simeq
  \{1.76,\ 1.98,\ 2.26,\ 2.49\}
  \quad
  \text{for}
  \quad
  a_{\rm latt}=\{1.0,\ 0.8,\ 0.6,\ 0.48\},
  \label{eq:tau_star_thermo_branches}
\end{equation}
with branch-fit residuals of order $10^{-3}$ in $\tau$. The key point is therefore robust: the dip time survives the $N\to\infty$ limit at fixed cutoff, so the spectral mechanism is not an artifact of keeping the physical volume fixed.
Panel (c) addresses the second limit in \eqref{eq:two_limits_tau}. The thermodynamic branch limits $\tau_\ast^{(\infty)}(a_{\rm latt})$ increase monotonically as the cutoff is reduced. A fit linear in $a_{\rm latt}$ gives
\begin{equation}
  \tau_\ast^{(\infty,0)}(m=-1.5)\simeq 3.12,
  \label{eq:tau_c_thermo_cont_linear}
\end{equation}
whereas a fit linear in $a_{\rm latt}^2$ gives $\tau_\ast^{(\infty,0)}\simeq 2.63$ and a visibly larger residual. 
The spread between the two extrapolation forms should be regarded as a systematic uncertainty of the continuum estimate.
At the current resolution, the safe statement is therefore not a precise intercept but the existence of a continued drift to later times, with the combined thermodynamic/continuum crossing lying at $\tau = O(3)$ and the empirical linear-in-$a_{\rm latt}$ fit favoring $\tau_\ast^{(\infty,0)} \approx 3.1$.
By contrast, panel (d) shows that the thermodynamic extrapolation of the dip depth $\Delta_\ast^{(\infty)}(a_{\rm latt})$ is much less regular; one branch even yields a slightly negative intercept. We do not observe a controlled scaling of the minimum gap itself in the thermodynamic/continuum limit at the present resolution, and therefore do not claim a sharp gap closure. The robust quantity at the present level is the location of the late-time crossing, not the detailed scaling of the minimum gap itself.

This two-step analysis sharpens the interpretation of \Cref{sec:specdyn}. The moving narrow-gap region identified in exact diagonalization survives the fixed-cutoff thermodynamic limit and continues to drift to later times as the lattice spacing is decreased. What is presently controlled is the persistence and location of the crossing time; the asymptotic sharpness of the dip still requires finer lattices and larger volumes. Even so, \Cref{fig:largeN_fixedmass} provides a concrete large-$N$ spectral anchor for the late-time window in which the corresponding mixed-state irreversibility front should remain visible.

\section{Entropy production and operational irreversibility front}
\label{sec:front}

We now switch to the Gibbs-state protocol and use the relative entropy \eqref{eq:sigma} to quantify how far the actual state is from the instantaneous thermal manifold at the same inverse temperature $\beta$. This reformulation turns the expansion-driven sweep into an explicitly operational nonequilibrium problem.

\subsection{Relative entropy as irreversible work and front theory}
Because the initial state is the Gibbs state of $H(0)$ and the subsequent dynamics is unitary, the relative entropy \eqref{eq:sigma} admits the exact rewriting
\begin{equation}
  \Sigma(\tau,m;\beta)
  = \beta\!\left[ W(\tau,m)-\Delta F_\beta(\tau,m)\right],
  \label{eq:sigma_irrev_work}
\end{equation}
where
\begin{equation}
  W(\tau,m)=\Tr[\rho(\tau)H(\tau)]-\Tr[\rho_0 H(0)],
  \qquad
  \Delta F_\beta(\tau,m)=F_\beta(H(\tau))-F_\beta(H(0)),
\end{equation}
and $F_\beta(H)=-\beta^{-1}\log\Tr e^{-\beta H}$. Thus the front is not merely a geometric feature of state space: it is the locus where the expansion-generated irreversible work becomes large compared with the equilibrium free-energy change.

To connect this directly with the narrow-gap picture, one may project onto the dominant two-level sector \eqref{eq:H_eff_AB} and then dephase in the instantaneous energy basis. By monotonicity of relative entropy under this projective measurement,
\begin{equation}
  \Sigma(\tau,m;\beta)\ge
  D_{\rm cl}\!\left(
  \{1-p_{\rm ex}^{AB}(\tau,m),\,p_{\rm ex}^{AB}(\tau,m)\}
  \,\Big\Vert\,
  \{1-p_{\beta}^{\rm eq}(\tau,m),\,p_{\beta}^{\rm eq}(\tau,m)\}
  \right),
  \label{eq:sigma_2level_bound}
\end{equation}
where $p_{\rm ex}^{AB}$ is the actual occupation of the upper adiabatic branch and
\begin{equation}
  p_{\beta}^{\rm eq}(\tau,m)=\frac{1}{1+\exp[\beta\Delta_{AB}(\tau,m)]}.
  \label{eq:peq_2level}
\end{equation}
The lower bound \eqref{eq:sigma_2level_bound} already explains the observed front: $p_{\rm ex}^{AB}$ rises rapidly when the sweep passes through the avoided crossing, whereas $p_{\beta}^{\rm eq}$ remains controlled by the instantaneous gap. The steepest mismatch therefore occurs near $\delta_{AB}(\tau,m)=0$, namely near the pseudo-critical line.

Using \eqref{eq:gap_eff}, one finds that the mass scale over which the equilibrium occupation changes is set by
\begin{equation}
  \delta m_T^{AB}\sim \frac{1}{\beta\abs{\Delta M^{AB}}},
\end{equation}
while the intrinsic avoided-crossing width is $\delta m_{\rm valley}^{AB}$ from \eqref{eq:valley_width_scaling}. Up to order-one coefficients, the front width is therefore controlled by
\begin{equation}
  \delta m_{\rm front}^{AB}
  \sim
  \max\!\left\{
  \frac{2\abs{\Gamma_{AB}(\tau)}}{\abs{\Delta M^{AB}}},
  \frac{1}{\beta\abs{\Delta M^{AB}}}
  \right\}.
  \label{eq:front_width_theory}
\end{equation}
Equation \eqref{eq:front_width_theory} gives a simple theoretical interpretation of the numerics: increasing $\beta$ removes thermal rounding, while increasing system size can sharpen the front whenever the finite-volume mixing $\Gamma_{AB}$ decreases or the relevant order-parameter contrast $\abs{\Delta M^{AB}}$ grows.

\subsection{Entropy-production landscape and front extraction}
Guided by \eqref{eq:sigma_irrev_work} and \eqref{eq:front_width_theory}, we now turn to the numerical extraction of the front. For visualization, it is convenient to display the rescaled quantity
\begin{equation}
  \widetilde\Sigma(\tau,m)=\log_{10}\!\left[1+\frac{\Sigma(\tau,m)}{\beta}\right].
  \label{eq:sigma_tilde}
\end{equation}
At low temperature, the resulting heat map develops a sharp boundary separating a low-irreversibility region from a high-irreversibility region; see \Cref{fig:dpt_summary}(a). To extract this boundary quantitatively, we shift each time slice by the pseudo-critical mass,
\begin{equation}
  x = m-\mc(\tau),
\end{equation}
and fit the profile to the sigmoid form
\begin{equation}
  \widetilde\Sigma(\tau,x)\simeq y_0(\tau)+\frac{A(\tau)}{1+\exp[-(x-x_f(\tau))/w(\tau)]}.
  \label{eq:sigmoid}
\end{equation}
The fitted center defines the operational dynamical critical mass
\begin{equation}
  \mDPT(\tau,\beta)=\mc(\tau)+x_f(\tau,\beta),
  \label{eq:mDPT}
\end{equation}
while the 25--75 crossing width is
\begin{equation}
  \Delta m_{25\to 75}(\tau)=2\ln 3\,w(\tau).
  \label{eq:width_def}
\end{equation}
In practice we quote widths only after the front becomes statistically detectable, defining an onset time $\tau_{\rm onset}$ from the fit quality and amplitude.

The central physical observation is immediate from \Cref{fig:dpt_summary}(a),(b): the front extracted from entropy production closely tracks the pseudo-critical line obtained from the instantaneous gap. The nonequilibrium front is therefore not an arbitrary contour in a noisy heat map. It is the operational image of the same moving narrow-gap valley already identified in \Cref{sec:specdyn}.

\subsection{Temperature sharpening and finite-size sharpening}
The temperature dependence of the front is summarized in \Cref{fig:dpt_summary}(b) and (c). As the inverse temperature increases, the post-onset front width narrows and the extracted line remains tied to the pseudo-critical trajectory, although finite-volume offsets from $\mc(\tau)$ are not strictly monotonic for every value of $\beta$. In other words, cooling the initial state does not merely rescale the heat map; it sharpens the irreversibility threshold itself. This is precisely what one expects if the front is controlled by a competition between adiabatic following and the moving many-body narrow-gap region.

The same interpretation is strengthened by the finite-size trends in \Cref{fig:dpt_summary}(e). At fixed low temperature, increasing $N$ decreases the mean front width, decreases the mean offset $\langle|\mDPT-\mc|\rangle$, and increases the fit-based sharpness. The data do not yet prove a singular thermodynamic-limit transition, but they do show the monotonic finite-size sharpening expected of a finite-volume precursor of dynamical criticality.

\subsection{LOCC witnesses and front reconstruction}
A key advantage of the relative-entropy formulation is that it can be restricted to local data. Two distant observers, Alice and Bob, each controlling an end block of size $\ell$, need not reconstruct the full many-body density matrix. They may instead use the classical measurement witness $\Sigmeas$ or the reduced-state witness $\Sigtomo$ defined in \eqref{eq:locc_witnesses}. Because both are lower bounds on the full relative entropy, they inherit a direct operational meaning. The ordering is structural, not merely numerical: $\Sigtomo$ stops after the partial trace onto the accessible blocks, whereas $\Sigmeas$ applies an additional local measurement channel, so tomography can fail to outperform direct measurement only when the chosen measurement basis already captures all of the reduced-state distinguishability.

\begin{figure*}[t]
  \centering
  \includegraphics[width=\textwidth]{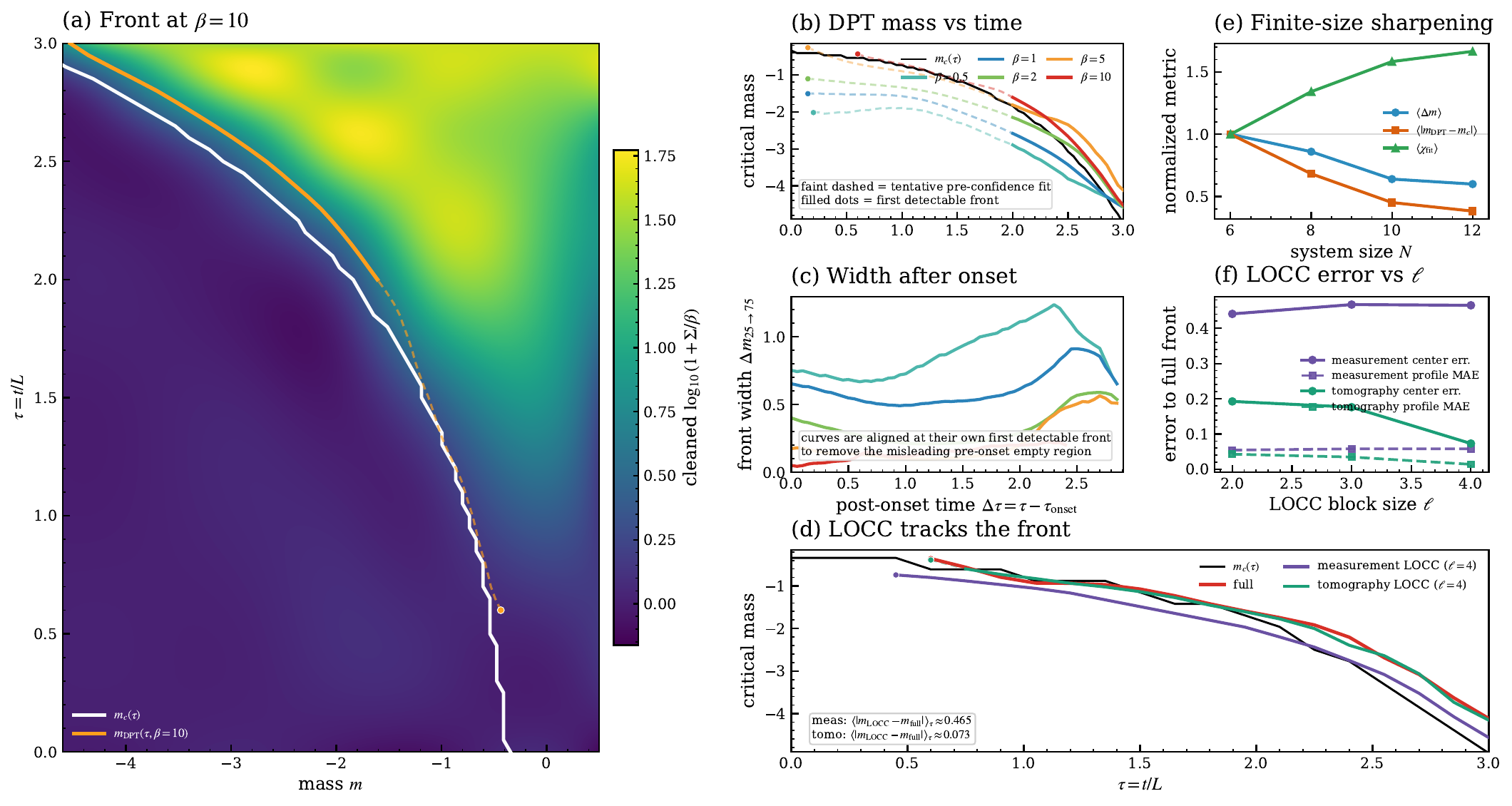}
  \caption{\textbf{Operational irreversibility front and its LOCC signatures.}
  \textbf{(a)} Smoothed entropy-production landscape $\widetilde\Sigma(\tau,m)$ at $\beta=10$ for a representative $N=10$ run. The white curve is the pseudo-critical line $\mc(\tau)$ and the orange curve is the extracted operational front $\mDPT(\tau,\beta=10)$. \textbf{(b)} Temperature dependence of the extracted front for $\beta=0.5,1,2,5,10$ together with $\mc(\tau)$. \textbf{(c)} Front width after onset, showing systematic sharpening at larger $\beta$. \textbf{(d)} Comparison between the full front and two LOCC reconstructions for end blocks of fixed size. \textbf{(e)} Finite-size sharpening of the front at low temperature: the mean width and the mean offset to $\mc(\tau)$ decrease, while the fit-based sharpness increases with $N$. \textbf{(f)} Error to the full front versus LOCC block size $\ell$, showing that tomography-based LOCC outperforms direct local measurement and improves strongly with $\ell$, while the direct measurement witness is much less sensitive to $\ell$ for the chosen measurement basis.}
  \label{fig:dpt_summary}
\end{figure*}

The comparison is shown in \Cref{fig:dpt_summary}(d) and (f). The measurement-based LOCC protocol follows the overall front trajectory but exhibits a noticeable displacement from the full result. The tomography-based protocol is significantly more faithful for the same block size, showing that the front is already encoded in the reduced state on the accessible subsystems. Increasing $\ell$ reduces the tomography-based error markedly, whereas the direct measurement witness is comparatively insensitive to $\ell$ for the present basis choice and remains displaced from the full front. Thus additional local access is most useful when the reduced-state information is retained rather than immediately projected onto the charge basis. This establishes that the irreversibility front is not merely a global diagnostic available only to an all-knowing observer. It is a feature that can be inferred by distant parties with limited but increasing local access.

\section{Discussion}
\label{sec:discussion}

The point to stress is that the novelty here is not simply that the Schwinger model can be placed in an expanding background. Recent work has already shown that interacting matter fields, and very recently the Schwinger model itself, can be formulated and evolved nonperturbatively in curved spacetime \cite{FulgadoClaudio2024,BuddEtAl2026}. What is specific to the present paper is the identification of a de Sitter-driven spectral mechanism in an interacting gauge theory: expansion suppresses the kinetic term, enhances the electric-energy scale, and thereby sweeps the many-body spectrum through a moving avoided-crossing structure. In the diagonal/off-diagonal decomposition of \Cref{sec:diabatic_crossing}, a family of configuration-space crossings drifts with $a(\tau)$ while the hopping turns each crossing into an avoided crossing of width $O(a^{-1})$. This deforms the instantaneous spectrum, shifts the valley of minimal gap to more negative mass, and drives the state through a moving near-critical region. Ground-state diagnostics reveal the coherent consequences of that sweep---loss of adiabatic following, excitation growth, and redshifted response---while Gibbs-state diagnostics reveal its irreversible consequence through the relative entropy to the instantaneous thermal manifold.

The nontriviality emerges from the quantitative entanglement of spectral and thermodynamic descriptions. The pseudo-critical line extracted from the spectrum and the front extracted from entropy production track each other closely. Lower temperature and larger system size both sharpen the front and reduce its offset from the gap valley. This is why we use the language of ``dynamical criticality'' rather than a generic crossover: the evidence points to a many-body threshold that becomes better defined as finite-size and thermal broadenings are reduced. At the same time, the present data remain finite-volume and the correct statement is therefore cautious. We view $\mDPT(\tau,\beta)$ as an operational dynamical critical mass and the observed front as the finite-volume precursor of a sharper transition that should be tested at larger $N$.

The fixed-mass MPS study of \Cref{sec:largeN} separates the two extrapolations that must not be conflated. The small-volume exact-diagonalization map is used to identify the moving narrow-gap mechanism across the full $(\tau,m)$ plane, but the thermodynamic statement comes from taking $\ell_{\rm phys}=N a_{\rm latt}\to\infty$ at each fixed cutoff. On every cutoff branch, increasing the physical volume and extrapolating in $1/\ell_{\rm phys}$ gives a stable dip time $\tau_\ast^{(\infty)}(a_{\rm latt})$, rising from about $1.76$ at $a_{\rm latt}=1$ to about $2.49$ at $a_{\rm latt}=0.48$. After first establishing the thermodynamic limit, we then examine how these branch limits drift as the cutoff is refined. The empirical linear-in-$a_{\rm latt}$ extrapolation favors a further shift to $\tau_\ast^{(\infty,0)}\approx 3.1$, whereas the $a_{\rm latt}^2$ check gives a smaller intercept and visibly larger residual. The safe conclusion is therefore twofold: the late-time crossing is not an artifact of fixed physical volume, and its location drifts to later times as the cutoff is reduced. By contrast, the dip depth remains too irregular for a universal scaling claim or for asserting a sharp continuum gap closure with the present data. In this sense, the finite-volume spectral map identifies the mechanism, while the sequential MPS extrapolation determines what survives the thermodynamic and continuum limits.

The LOCC analysis adds a genuinely quantum-informational layer to this interpretation. Relative entropy is not merely a formal distance to the instantaneous Gibbs state; it admits lower bounds that can be reconstructed from local data, and these lower bounds still carry the imprint of the irreversibility front. This suggests several directions for future work: shadow-tomography surrogates for the reduced-state witness, experimentally lighter measurement schemes, mixed-state tensor-network simulations at the larger sizes suggested by \Cref{sec:largeN}, and non-Abelian or Thirring \cite{THIRRING195891} (Gross-Neveu \cite{1974PhRvD..10.3235G})-type extensions where the competition between mass, redshift, and gauge dynamics is even richer. The framework is also naturally compatible with contemporary efforts to view lattice gauge theories as quantum-information platforms rather than only as discretized field theories \cite{Banuls2020}.

\section{Conclusion}
\label{sec:conclusion}
The central result of this paper is that de Sitter expansion in QED$_2$ does more than generate particle production in a curved background. Because the hopping redshifts as $1/a(t)$ while the electric term grows as $g^2 a(t)$, the expansion itself acts as a built-in drive across a moving narrow-gap region of the interacting spectrum. Starting from curved-space QED$_2$ in a spatially flat FLRW background, we derived the corresponding cosmic-time qubit Hamiltonian and used it to formulate an expansion-induced dynamical-criticality problem in a gauge theory with exact Gauss constraints.

Ground-state evolution shows that crossing the narrow-gap valley produces loss of adiabaticity, excitation energy growth, and a redshifted structure-factor response. The full spectral map used to identify this valley is displayed at fixed volume, where the $(\tau,m)$ structure can be resolved most clearly, but the survival of the late-time crossing is tested separately. Complementary matrix-product-state calculations at fixed $m=-1.5$ first take the fixed-cutoff thermodynamic limit, in which $\tau_\ast(a_{\rm latt},\ell_{\rm phys})$ converges as $\ell_{\rm phys}=N a_{\rm latt}\to\infty$, and then examine the continuum drift of the resulting branch limits. These thermodynamic dip times increase monotonically from $1.76$ to $2.49$ as $a_{\rm latt}$ is reduced from $1.0$ to $0.48$. An empirical linear extrapolation in $a_{\rm latt}$ favors $\tau_\ast^{(\infty,0)}\approx 3.1$, while the dip depth remains more cutoff sensitive and is not yet suitable for a universal scaling claim. Therefore the late-time crossing persists and moves to later times in the combined thermodynamic-continuum limit. Gibbs-state evolution shows that the same moving valley generates an entropy-production front, which defines an operational dynamical critical mass $\mDPT(\tau,\beta)$. The front tracks the pseudo-critical line, sharpens with inverse temperature, sharpens further with system size, and remains visible under LOCC, with tomography-based witnesses outperforming direct local measurement.

In aggregate, these results distinguish expanding QED$_2$ from both free-field particle-production problems and from purely spectral studies of lattice gauge theory. The new element is the quantitative bridge between a moving pseudo-critical structure in the instantaneous spectrum and an operational measure of irreversibility. The theoretical backbone is a drifting diabatic crossing $m_c^{AB}(\tau)$, produced by the growing electric term and a redshifted hopping matrix element that makes the corresponding avoided crossing narrower and more non-adiabatic at later times. What survives most clearly in the present data is the late-time crossing scale itself, rather than a fully established sharp continuum singularity in the gap depth. This makes QED$_2$ in de Sitter space a particularly sharp laboratory for future tensor-network studies and, eventually, for quantum-simulation experiments aimed at curved-spacetime gauge dynamics. A straightforward application of this work includes various interacting fermions and gauge theories in dS in AdS spaces \cite{Ikeda:2025lig,Ikeda:2025yqq}.
\section*{Data and code availability}
Source data and code for the work are available from authors' GitHub repository:\\
\url{https://github.com/IKEDAKAZUKI/Quantum-Information-Dynamics-in-de-Sitter-Space}

\section*{Acknowledgments}
The work of KI was supported by the U.S. Department of Energy, Office of Science, under Contract No.DE-SC0026415 and in part by the NSF under Grant No. OSI-2328774 in particular, on the connection between quantum fundamentals and quantum systems. The work of YO was supported by the Israeli Science Foundation Excellence Center, the US-Israel Binational Science Foundation, the Israel Ministry of Science.
\bibliographystyle{JHEP}
\bibliography{refs}

\end{document}